\documentclass{article}
\usepackage{arxiv,times}

\usepackage{amsmath,amsfonts,bm}

\def\eqref#1{equation~\ref{#1}}

\def\1{\bm{1}}

\DeclareMathAlphabet{\mathsfit}{\encodingdefault}{\sfdefault}{m}{sl}
\SetMathAlphabet{\mathsfit}{bold}{\encodingdefault}{\sfdefault}{bx}{n}

\usepackage{hyperref}
\usepackage{url}
\usepackage{booktabs}
\usepackage{longtable}
\usepackage{graphicx,xspace,enumitem,multirow,wrapfig,array}
\usepackage[most]{tcolorbox}

\newtcolorbox{finding}{colback=gray!12,colframe=gray!12,boxrule=0pt,
  arc=0pt,left=6pt,right=6pt,top=4pt,bottom=4pt,fontupper=\small}

\newcommand\ours{XRepoSkill\xspace}

\title{XRepoSkill: Learning Transferable Skills for Software Engineering Agents}

\author{%
Yaoqi Guo$^{1}$, Haoyang Zhou$^{2}$, Jiayi Zhang$^{1}$, Yiran Zhang$^{1}$, Yang Liu$^{1}$, \\
\bf Qiuyuan Chen$^{3}$, Qiang Lin$^{3}$, Hande Dong$^{3}$, Jie M. Zhang$^{4}$, Zhenpeng Chen$^{2}$\thanks{Corresponding author: Zhenpeng Chen.} \\[4pt]
\normalfont $^{1}$Nanyang Technological University \quad
$^{2}$Tsinghua University \\
$^{3}$Tencent \quad
$^{4}$King's College London \\[4pt]
\small\texttt{yaoqi001@e.ntu.edu.sg}, \texttt{zpchen@tsinghua.edu.cn}
}

\iclrfinalcopy
\begin{document}

\maketitle
\fancyhead{}                       
\renewcommand{\headrulewidth}{0pt} 

\begin{abstract}
Software engineering agents increasingly use reusable skills distilled from prior experience to resolve repository-level issues, yet such skills often fail to transfer across repositories. A central challenge is that a behavior appearing in a successful trajectory is not necessarily responsible for the successful outcome: it may be genuinely useful, merely incidental, or simply a recurring habit of the model. We introduce \ours, a trajectory-based approach for learning transferable skills. We represent a skill as a collection of rules, each specifying what action to take and when to take it during issue resolution. \ours first contrasts successful and failed trajectories of the same agent on the same issue and derives candidate rules from where their execution paths diverge. Each rule is paired with an executable predicate that enables its prescribed behavior to be evaluated systematically on other trajectories. A rule is verified based on its association with successful issue resolution and retained only when its prescribed behavior recurs across multiple repositories; repository-specific variants of the same behavior are then consolidated into transferable rules. For a new issue, \ours selects relevant rules to guide the agent. We learn skills from publicly released trajectories on the official SWE-bench Verified leaderboard and evaluate them on SWE-bench Pro and DeepSWE using three backbone LLMs from different vendors; none of the evaluation repositories appears in the skill-learning trajectory pool. Against three recent skill learning methods, \ours achieves the highest issue resolution rate in all six benchmark--LLM combinations. In particular, on the challenging long-horizon DeepSWE benchmark, \ours improves issue resolution by 10.3 percentage points over the same agent without learned skills and by 5.0 points over the strongest skill-learning baseline.
\end{abstract}

\section{Introduction}

Software engineering agents can resolve software issues by exploring a repository, modifying its code, running tests, and producing a patch~\citep{jimenez2024swebench,yang2024sweagent}. One increasingly popular way to equip such agents with reusable expertise, without retraining the backbone LLM, is through \emph{skills}~\citep{anthropic2026skills}: packages of task-specific knowledge and guidance loaded at inference time to steer agent behavior.

A central challenge, however, is that software engineering skills do not necessarily transfer across repositories. Recent work~\citep{han2026sweskillsbench} shows that most public software engineering skills provide little or no improvement on real-world repositories, and some can even hurt performance when their guidance does not fit the target context. This context dependence raises a fundamental question for skill learning: \emph{which behaviors learned from past repositories will remain useful when the agent encounters a repository it has never seen before?}

Recent work has begun to learn skills directly from agent trajectories, allowing agents to reuse experience from previous tasks. Trace2Skill~\citep{ni2026trace2skill}, for example, extracts trajectory-local lessons and hierarchically consolidates recurring ones into a shared skill directory; and, specifically for software engineering, STAIR~\citep{xu2026stair} abstracts past repair trajectories into reusable actions and strategies that are retrieved and adapted for new issues. Yet observing a behavior in a successful trajectory does not by itself establish that the behavior is relevant to the successful outcome: it may be genuinely useful, merely incidental, or simply reflect a recurring habit of the backbone LLM. Moreover, even a useful behavior may depend on repository-specific conventions or workflows and fail to transfer elsewhere. This makes it difficult to determine which behaviors extracted from historical trajectories constitute genuinely reusable software engineering knowledge.

To address this challenge, we propose \ours, a trajectory-based skill learning method that separates the discovery of potentially useful behaviors from their empirical verification. We represent a skill as a collection of rules, each specifying what action to take and when to take it during issue resolution. \ours first contrasts successful and failed trajectories produced by the same agent on the same issue with different backbone LLMs. Because the issue, repository, and agent are fixed, their shared behavior provides little evidence for explaining the different outcomes; \ours therefore localizes where the two execution paths first diverge and uses the subsequent behavioral differences to derive candidate rules. This comparison narrows rule discovery to behaviors that distinguish successful from unsuccessful attempts rather than indiscriminately summarizing everything that appears in a successful trajectory.

Crucially, candidate rules do not enter the skill simply because they were discovered from a successful attempt. \ours associates each rule with an executable predicate that determines whether a trajectory exhibits the prescribed behavior, enabling systematic evaluation on trajectories beyond the original pair. Each rule is verified based on the association between its prescribed behavior and issue-resolution outcomes. \ours then examines whether rules from different repositories capture the same underlying behavior and consolidates them into repository-independent transferable rules. The resulting rules form the learned skill. At inference time, \ours selects rules relevant to the current issue, and provides them as additional guidance to the software engineering agent.

We evaluate \ours on the widely used SWE-bench Pro benchmark~\citep{deng2025swebenchpro} and the more challenging long-horizon DeepSWE benchmark~\citep{datacurve2026deepswe} using three backbone LLMs from different vendors. For skill learning, \ours uses publicly released trajectories from the official SWE-bench Verified leaderboard~\citep{swebench2026leaderboard}. None of the repositories in either evaluation benchmark appears in this trajectory pool, ensuring that the evaluation directly measures transfer to previously unseen repositories. We use mini-SWE-agent~\citep{yang2024sweagent} as the underlying software engineering agent, a lightweight yet strong open-source framework that has been widely adopted as a research baseline and as the foundation for several recent software engineering agents~\citep{lee2026gistify,guo2026swe,tripathy2026swenergy}. We compare \ours with three recently proposed skill learning methods under the same agent and trajectory setting.

Across all six benchmark--LLM combinations, \ours consistently achieves the highest issue resolution rate. In particular, on the more challenging long-horizon DeepSWE benchmark, \ours improves issue resolution by 10.3 percentage points over the same agent without learned skills and by 5.0 points over the strongest skill learning baseline. The learned skill also transfers beyond the programming language represented during skill learning: although all source trajectories come from Python repositories, \ours achieves clear improvements on issues written in other programming languages. These results indicate that \ours distills transferable issue-resolution guidance rather than repository- or language-specific patterns.

In summary, this paper makes the following contributions:
\begin{itemize}[leftmargin=*]

\item We introduce \ours{}, a trajectory-based skill learning approach that discovers candidate rules from behavioral divergences between successful and failed attempts, verifies them on independent trajectory evidence, and retains only behaviors recurring across repositories as transferable rules.

\item We extensively evaluate \ours against recent skill learning methods on SWE-bench Pro and DeepSWE, showing the highest issue resolution rate in all evaluated benchmark--LLM combinations and strong transfer to previously unseen repositories and programming languages.
\item We publicly release our code and data at \url{https://github.com/XREPOSKILL/XRepoSkill}.
\end{itemize}

\section{Methodology}
\label{sec:method}

\subsection{Problem Formulation}
\label{sec:method:problem}
We consider a software engineering agent $A$ and a trajectory pool $\mathcal{T}$ that records its attempts to resolve historical software issues using different backbone LLMs. Each trajectory contains the agent's sequence of actions and observations during one issue-resolution attempt, and $\mathcal{T}$ includes both successful and failed trajectories.

From $\mathcal{T}$, we seek to learn a transferable skill $S$. $S$ consists of a collection of rules, each providing reusable guidance on what action to take and when to take it during issue resolution. We intend the learned skill to support future issues from repositories that do not appear in the trajectory pool.

For a new issue $i^\star$, \ours selects relevant rules from $S$ and provides them to $A$ as additional guidance. Our goal is for the resulting skill-guided agent to achieve a higher issue resolution rate on previously unseen repositories than the original agent without the learned skill.

\subsection{\ours{}: In a Nutshell}
\label{sec:method:overview}
\begin{figure}[t]
\centering
\includegraphics[width=0.95\linewidth]{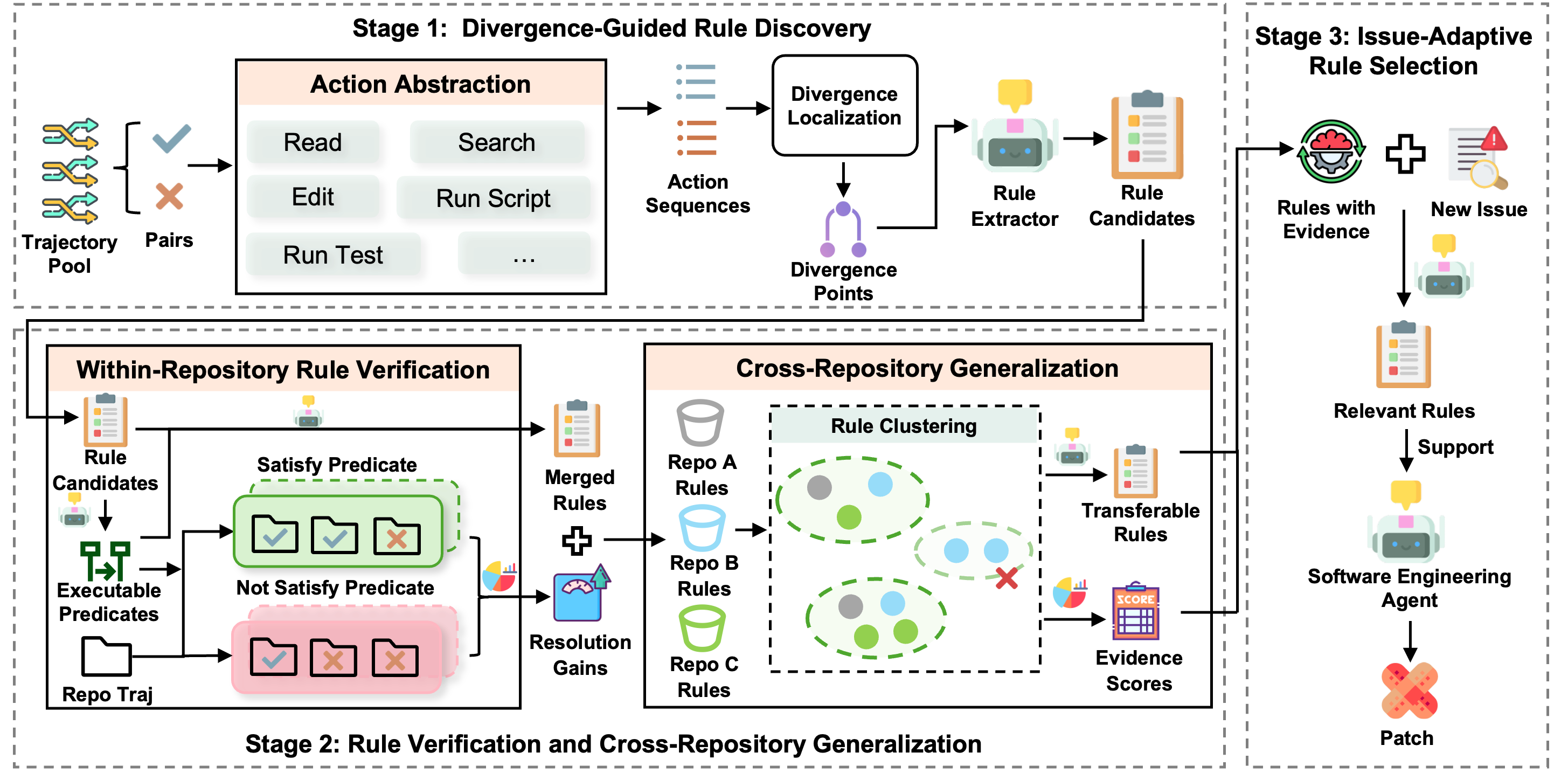}
\caption{Overview of \ours.}
\label{fig:overview}
\end{figure}

Figure~\ref{fig:overview} presents an overview of \ours. Given the trajectory pool $\mathcal{T}$ of the software engineering agent $A$, \ours distills a transferable skill $S$ and applies it to a new issue $i^\star$. \ours consists of three stages. First, \textbf{Divergence-Guided Rule Discovery} (Section~\ref{sec:method:fork}) compares successful and failed trajectories of $A$ on the same issue, identifies where their execution paths first diverge, and turns the divergent actions into candidate rules. Second, \textbf{Rule Verification and Cross-Repository Generalization} (Section~\ref{sec:method:distill}) evaluates each candidate rule on other trajectories to determine whether trajectories that perform its prescribed action resolve issues more often than those that do not. It then consolidates rules from different repositories that capture the same underlying behavior into transferable rules, forming the skill $S$. Third, \textbf{Issue-Adaptive Rule Selection} (Section~\ref{sec:method:inject}) selects rules relevant to $i^\star$ from $S$ and provides them as guidance to the agent $A$.

\subsection{Stage 1: Divergence-Guided Rule Discovery}\label{sec:method:fork}
The goal of this stage is to discover candidate rules from behavioral differences between successful and failed attempts on the same issue. For each issue in $\mathcal{T}$ with both outcomes, \ours pairs a successful trajectory with a failed trajectory produced by the same agent $A$ under different backbone LLMs. Because the issue, repository, and agent are fixed, behaviors shared by the two trajectories cannot explain their different outcomes. \ours therefore aligns the trajectories, identifies where their execution paths first meaningfully diverge, and uses the subsequent behavioral differences to derive candidate rules. At this stage, these rules are only hypotheses; whether they are consistently associated with successful issue resolution is examined in Stage~2.

\paragraph{Action abstraction.} It is unreliable to directly compare raw trajectories because the same behavior may be expressed through different tool calls. Thus, \ours maps each tool call to an abstract action defined by its \emph{type} (e.g., reading a file, searching, editing, or running tests) and \emph{target} (i.e., the object on which the call operates). For example, different commands that read the same file are mapped to the same abstract action. Each trajectory is thereby transformed into an action sequence, allowing behaviorally equivalent steps to be aligned despite differences in their concrete realization.

\paragraph{Divergence localization.} \ours aligns the two action sequences from the beginning and identifies the first point at which their behaviors meaningfully differ. Two actions match when they have the same type and target. Differences in the ordering of information-gathering actions, such as reading and searching, are tolerated because they may reflect different exploration orders rather than different resolution strategies. In contrast, an unmatched action that materially changes the subsequent course of the attempt, such as editing code or running a test, terminates the shared prefix. We define the first action after this shared prefix as the divergence point. The divergence point serves as an anchor for rule discovery: actions before it are shared by both the successful and failed attempts and therefore provide little evidence for explaining their different outcomes, whereas the behavior after it contains decisions unique to one execution path and thus provides a more informative search space for candidate rules. \ours retains the shared prefix in condensed form together with the two trajectory suffixes beginning at the divergence point for rule generation.

\paragraph{Candidate rule generation.} The first divergence is not necessarily the behavior most relevant to the final outcome: once the trajectories separate, useful actions may occur later in the successful execution path. Rather than comparing the two suffixes step by step after the divergence point, \ours extracts candidate rules from three complementary reference points that remain well-defined without step-by-step alignment. The first examines the failed trajectory in light of the successful patch to identify information that may have helped the failed attempt. The second examines the successful trajectory to identify useful information-gathering and validation performed before editing. The third focuses on the divergence point itself and asks what the successful attempt does that the failed one does not. Each identified behavior is converted into a candidate rule specifying both \emph{what action to take} and \emph{when to take it}, and is phrased to apply beyond the specific issue from which it was extracted. The resulting candidate rules are passed to Stage~2.

\subsection{Stage 2: Rule Verification and Cross-Repository Generalization}\label{sec:method:distill}

Stage~1 produces candidate rules from individual trajectory pairs, but a behavior that distinguishes one pair may be incidental to that pair or specific to its repository. Stage~2 therefore asks two questions: to what extent the behavior prescribed by a candidate rule is associated with successful issue resolution beyond the pair from which it is discovered, and whether the same behavior recurs across repositories. Rules capturing the same behavior across repositories are consolidated into the transferable skill $S$.

\paragraph{Executable rule predicates.} Candidate rules are written in natural language, making repeated evaluation over extensive trajectories with LLM calls both expensive and potentially inconsistent. During candidate rule generation, \ours therefore produces an executable predicate together with each rule. The predicate is defined over the abstract actions from Stage~1 and captures the behavior prescribed by the rule, including whether an action occurs, whether one action precedes another, or how often an action occurs. For instance, the rule ``run the failing test before editing any file'' is represented by the predicate $\mathrm{before}(\mathrm{run\_test},\mathrm{first\_edit})$. Evaluating the predicate on a trajectory requires no LLM call and deterministically indicates whether the trajectory exhibits the behavior described by the rule. Each rule thus partitions the trajectory pool $\mathcal{T}$ into trajectories that satisfy its predicate and those that do not, providing the basis for the verification that follows.

\paragraph{Within-repository rule verification.} Different trajectory pairs from the same repository may yield candidate rules that describe the same behavior in different words. \ours therefore first merges, within each repository, candidate rules that describe the same behavior into one rule, using an LLM to consolidate their descriptions and select the predicate that best captures the shared behavior. \ours then verifies each merged rule by comparing the outcomes of trajectories that satisfy its predicate with those that do not. To avoid circular validation, this comparison excludes the successful--failed pairs from which any constituent rule was derived; otherwise, the rule could appear useful simply because it was extracted from a pair in which the successful trajectory exhibited the behavior and the failed trajectory did not.

The comparison is performed within issues rather than by pooling trajectories across issues. This controls for issue difficulty: a behavior may occur more often on easy issues and thus appear correlated with success even when it provides no benefit. For issue $j$, let $n_{1j}$ and $n_{0j}$ denote the numbers of trajectories that satisfy and do not satisfy the rule predicate, and let $a_j$ and $c_j$ denote the corresponding numbers of successful trajectories. We define the rule's \emph{resolution gain} as
\begin{equation}
\mathrm{gain}
=
\frac{\sum_j w_j \left(a_j/n_{1j} - c_j/n_{0j}\right)}
{\sum_j w_j},
\qquad
w_j
=
\frac{n_{1j}n_{0j}}{n_{1j}+n_{0j}},
\label{eq:gain}
\end{equation}
which is the Mantel--Haenszel estimator of a difference in rates across strata~\citep{mantel1959statistical}. An issue contributes only when both groups are represented among its trajectories; otherwise, it provides no evidence for comparing them. Each rule also receives a $z$ score, defined as its gain divided by its standard error~\citep{sato1989variance}, so that the strength of the evidence reflects both the estimated gain and the amount of supporting data. The gain and $z$ score are not used as thresholds for retaining rules; instead, they inform issue-adaptive rule selection in Stage~3.

\paragraph{Cross-repository generalization.}
Evidence from a single repository is insufficient to establish transferability: a rule may depend on repository-specific conventions, workflows, or commands. \ours{} therefore requires the same behavior to be discovered and independently verified in multiple repositories before it enters the transferable skill.

To identify rules that capture the same behavior across repositories, \ours compares their executable predicates rather than their natural-language wording. Each predicate is evaluated on every trajectory in $\mathcal{T}$, turning each rule into a binary vector with one entry per trajectory that indicates whether the trajectory satisfies the predicate. Rules that prescribe the same behavior should therefore produce similar vectors. We measure this similarity using the $\phi$ correlation \citep{yule1912methods} between two binary vectors rather than raw agreement: because most predicates are false on most trajectories, two unrelated rules may agree frequently simply because both are absent, whereas correlation discounts such shared absence.

\ours groups rules using agglomerative clustering with distance $1-\phi$ and complete linkage \citep{defays1977efficient}. This requires each pair of rules within a cluster to be sufficiently similar, preventing unrelated rules from being grouped together only through chains of intermediate rules. A cluster is retained only if it contains independently verified rules from at least two repositories. Its members are then consolidated by an LLM into a \textbf{transferable rule}, rewritten to remove repository-specific paths, symbols, and other identifiers. Clusters supported by only one repository are excluded.

A transferable rule may be supported with different strength across repositories. Let $g_k$ and $s_k$ denote the gain and standard error of the rule instance from repository $k$. \ours{} combines these estimates using inverse-variance weighting:
\begin{equation}
\bar{g}
=
\frac{\sum_k g_k/s_k^2}{\sum_k 1/s_k^2},
\qquad
\bar{s}
=
\left(\sum_k 1/s_k^2\right)^{-1/2},
\label{eq:pool}
\end{equation}
so that more precise estimates receive greater weight~\citep{cochran1954combination}. The transferable rule receives gain $\bar{g}$ and $z$ score $\bar{g}/\bar{s}$. The resulting skill $S$ consists of these transferable rules together with their evidence scores, which are used in Stage~3 for issue-adaptive rule selection.

\subsection{Stage 3: Issue-Adaptive Rule Selection}\label{sec:method:inject}
Stage~3 applies the learned skill $S$ to a new issue $i^\star$ without modifying the underlying agent $A$. \ours changes only the additional guidance provided for the issue.

\paragraph{Rule selection.} Given $i^\star$ and $S$, \ours asks the backbone LLM of $A$ to select rules relevant to the issue. This selection is necessary because not every rule applies to every issue, and providing too many rules may dilute useful guidance. During selection, each rule is accompanied by its evidence, i.e., its gain and $z$ score, which summarize the historical evidence supporting the rule and help the selector distinguish broadly supported guidance from weaker candidates.

\paragraph{Skill-guided resolution.} The selected rules are provided to $A$ as issue-specific guidance. The agent is instructed to verify each suggested action against the current repository before applying it, so the rules guide rather than override its reasoning over the issue and codebase. It then resolves $i^\star$ using its original tools and execution procedure, with the selected rules as the only additional guidance introduced by \ours.

\section{Evaluation}
\label{sec:eval}

\subsection{Research Questions (RQs)}

\noindent \textbf{RQ1 (Effectiveness):} How effective is \ours at improving issue resolution on previously unseen repositories?

\noindent \textbf{RQ2 (Cross-Language Generalization):} Can skills distilled by \ours generalize to issues in different programming languages?

\noindent \textbf{RQ3 (Stability):} How stable is the effectiveness of \ours across repeated runs?

\noindent \textbf{RQ4 (Ablation Study):} How do \ours's key design choices affect its effectiveness?

\subsection{Datasets and Evaluation Metric}

\paragraph{Trajectory pool used for skill learning.}
We use publicly released trajectories from the official SWE-bench Verified leaderboard~\citep{swebench2026leaderboard} as the source data for skill learning. These trajectories are generated by running mini-SWE-agent 2.0 (hereafter, mini-SWE-agent) on SWE-bench Verified~\citep{jimenez2024swebench}, a widely used issue-resolution benchmark comprising 500 real-world, human-validated Python issues. mini-SWE-agent is a lightweight yet strong open-source agent that is widely adopted as a research baseline \citep{lee2026gistify,guo2026swe,tripathy2026swenergy}. The leaderboard includes trajectories produced with 12 backbone LLMs: Claude Haiku 4.5, Claude Sonnet 4.5, Claude Opus 4.5, Claude Opus 4.6, DeepSeek-V3.2, Gemini 3 Flash, Gemini 3 Pro, GLM-5, GPT-5.2, GPT-5 mini, Kimi K2.5, and MiniMax M2.5. Together, these runs yield a trajectory pool with 12 pass- or fail-labeled trajectories per issue.

\paragraph{Evaluation Benchmarks.}
We evaluate \ours on two recently introduced benchmarks, SWE-bench Pro \citep{deng2025swebenchpro} and DeepSWE \citep{datacurve2026deepswe}.
\begin{itemize}[leftmargin=*]
\item \textbf{SWE-bench Pro} contains 731 issue resolution tasks from 11 open-source repositories spanning Python, Go, JavaScript, and TypeScript. Due to budget constraints, we randomly sample 200 tasks for evaluation, comprising 69 Python, 73 Go, 52 JavaScript, and 6 TypeScript tasks. \item \textbf{DeepSWE} contains 113 long-horizon issue resolution tasks specifically designed for the benchmark rather than mined from public issue trackers, including 34 Python, 34 Go, 5 JavaScript, 35 TypeScript, and 5 Rust tasks. We evaluate on all 113 tasks.
\end{itemize}
We have confirmed that none of the repositories in either benchmark appears in our trajectory pool, ensuring that the evaluation measures transfer to previously unseen repositories.

\paragraph{Evaluation Metric.}
We use the resolution rate (\textbf{\%Resolved}) as the evaluation metric, defined as the proportion of benchmark issues resolved by an agent.

\subsection{Baselines and LLMs}
Besides the raw mini-SWE-agent without any skills, we compare \ours with three representative trajectory-based skill-learning methods:

\begin{itemize}[leftmargin=*]
\item \textbf{SkillOpt} \citep{yang2026skillopt} treats the skill document as a trainable state of an agent and iteratively updates it from scored trajectories, retaining only changes that improve held-out performance.
\item \textbf{STAIR} \citep{xu2026stair} abstracts past repair trajectories into reusable actions and strategies, then retrieves and adapts relevant experience into an issue-specific plan for each new task.
\item \textbf{Trace2Skill} \citep{ni2026trace2skill} distills trajectory-local lessons in parallel and hierarchically consolidates them into a shared transferable skill directory used at inference time.
\end{itemize}

To ensure a fair comparison, we use the same LLM, DeepSeek-V4-Pro-0813, for \ours and all baseline methods to learn skills from the same trajectory pool.

For issue resolution, we run all methods with the same mini-SWE-agent and evaluate each with three backbone LLMs from different vendors: DeepSeek-V4-Flash-0731, GPT-5.6-Luna, and MiMo-V2.5-Pro, hereafter referred to as \textbf{DeepSeek}, \textbf{GPT}, and \textbf{MiMo}. We set the reasoning effort to high for all three models, as issue resolution requires reasoning over the issue description, repository context, and tool outputs before making code edits.

\section{Results and Analysis}

\subsection{RQ1: Effectiveness}
\label{sec:main}

RQ1 evaluates the effectiveness of \ours in resolving issues from previously unseen repositories. Table~\ref{tab:pass_rate} reports the resolution rates of \ours and the baselines across three backbone LLMs on SWE-bench Pro and DeepSWE. \ours achieves the highest resolution rate in all six benchmark--LLM combinations. This consistency is notable because the strongest baseline varies across benchmarks: on average, STAIR performs best on SWE-bench Pro, whereas Trace2Skill is strongest on DeepSWE. In contrast, \ours remains the top-performing method across both benchmarks and all three LLMs, demonstrating more consistent effectiveness than existing skill-learning methods.

On SWE-bench Pro, \ours achieves an average resolution rate of 53.8\%, outperforming mini-SWE-agent and the strongest baseline, STAIR, by 6.0 and 2.1 percentage points, respectively. The margin is larger on the more challenging long-horizon DeepSWE, where \ours reaches 44.2\%, exceeding mini-SWE-agent and the strongest baseline, Trace2Skill, by 10.3 and 5.0 points, respectively. This larger gain suggests that \ours is particularly effective on more demanding tasks.

\begin{table}[t]
\centering
\caption{(RQ1) Issue resolution rates (\%) of different methods across LLMs on SWE-bench Pro and DeepSWE. Avg. denotes the mean over the three LLMs for each benchmark. The highest rate in each column is shown in bold.}
\label{tab:pass_rate}
\setlength{\tabcolsep}{4pt}
\small
\begin{tabular}{lrrrrrrrr}
\toprule
& \multicolumn{4}{c}{SWE-bench Pro}
& \multicolumn{4}{c}{DeepSWE} \\
\cmidrule(lr){2-5}\cmidrule(lr){6-9}
Method
& \multicolumn{1}{c}{DeepSeek}
& \multicolumn{1}{c}{GPT}
& \multicolumn{1}{c}{MiMo}
& \multicolumn{1}{c}{Avg.}
& \multicolumn{1}{c}{DeepSeek}
& \multicolumn{1}{c}{GPT}
& \multicolumn{1}{c}{MiMo}
& \multicolumn{1}{c}{Avg.} \\
\midrule
mini-SWE-agent & 51.5 & 50.5 & 41.5 & 47.8 & 46.0 & 44.2 & 11.5 & 33.9 \\
\midrule
SkillOpt       & 54.0 & 51.5 & 41.0 & 48.8 & 40.7 & 39.8 & 13.3 & 31.3 \\
STAIR          & \textbf{55.5} & 51.0 & 48.5 & 51.7 & 43.4 & 46.0 & 14.2 & 34.5 \\
Trace2Skill    & 55.0 & 52.0 & 45.0 & 50.7 & 49.6 & 53.1 & 15.0 & 39.2 \\
\midrule
\ours      & \textbf{55.5} & \textbf{56.0} & \textbf{50.0} & \textbf{53.8}
               & \textbf{58.4} & \textbf{56.6} & \textbf{17.7} & \textbf{44.2} \\
\bottomrule
\end{tabular}
\end{table}

We also compare the runtime cost of each method. The costs are broadly comparable across methods. On SWE-bench Pro, \ours costs \$0.11 per issue on average, close to mini-SWE-agent (\$0.09), SkillOpt (\$0.12), STAIR (\$0.10), and Trace2Skill (\$0.10). On DeepSWE, \ours costs \$0.25 per issue, compared with \$0.22 for mini-SWE-agent, SkillOpt, and STAIR, and \$0.18 for Trace2Skill. Thus, \ours operates at a similar cost scale to existing methods while achieving the highest resolution rate in all six benchmark--backbone combinations, indicating a favorable effectiveness--cost trade-off.

\begin{table}[t]
\centering
\caption{(RQ2) Resolution-rate gains (\%) of \ours over STAIR and Trace2Skill across programming languages, averaged over three backbone LLMs. \# denotes the number of issues for each language in each subset.}
\label{tab:language}
\small
\setlength{\tabcolsep}{5pt}
\begin{tabular}{lrrr@{\hskip 12pt}rrr}
\toprule
\multirow{2}{*}{Language}& \multicolumn{3}{c}{SWE-bench Pro} & \multicolumn{3}{c}{DeepSWE} \\
\cmidrule(lr){2-4}\cmidrule(lr){5-7}
 & \# & vs. STAIR & vs. Trace2Skill & \# & vs. STAIR & vs. Trace2Skill \\
\midrule
Python     & 69 & $-1.0$ & $+1.0$ & 34 & $+5.9$  & $+0.0$  \\
Go         & 73 & $+2.7$ & $+4.1$ & 34 & $+2.0$  & $+3.9$  \\
JavaScript & 52 & $+6.4$ & $+4.5$ & 5  & $+13.3$ & $+0.0$  \\
TypeScript & 6  & $-5.6$ & $+5.6$ & 35 & $+21.0$ & $+10.5$ \\
Rust       & -- & --     & --     & 5  & $+6.7$  & $+13.3$ \\
\bottomrule
\end{tabular}
\end{table}

\subsection{RQ2: Cross-Language Generalization}
\label{sec:language}
RQ2 examines whether the skills learned by \ours from Python-only repositories generalize to other programming languages. Since RQ1 identifies STAIR and Trace2Skill as the strongest baselines on SWE-bench Pro and DeepSWE, respectively, Table~\ref{tab:language} compares \ours with both and reports the resolution-rate gains for each programming language, averaged over the three backbone LLMs. The results show that \ours remains competitive on Python while achieving clearer advantages on languages absent from the skill-learning corpus, suggesting that its learned rules transfer beyond the language from which they are distilled.

On Python, the only language represented in the trajectory pool, \ours performs similarly to the strongest baselines. On SWE-bench Pro, it trails STAIR by 1.0 percentage point and exceeds Trace2Skill by 1.0 point; on DeepSWE, it outperforms STAIR by 5.9 points and matches Trace2Skill. The advantage becomes more consistent once the programming language changes. On Go, \ours outperforms both baselines on both benchmarks, by 2.0--4.1 points. On JavaScript in SWE-bench Pro, it exceeds STAIR and Trace2Skill by 6.4 and 4.5 points, respectively. On TypeScript in DeepSWE, the gains further increase to 21.0 and 10.5 points. Together, these results suggest that \ours distills issue-resolution procedures that generalize across programming languages rather than relying primarily on Python-specific patterns.
The results for TypeScript on SWE-bench Pro and for JavaScript and Rust on DeepSWE should be interpreted with caution, as these subsets contain only 6, 5, and 5 issues, respectively.

\subsection{RQ3: Stability}\label{sec:stability}
\begin{wraptable}{r}{0.45\textwidth}
\vspace{-12pt}
\centering
\caption{(RQ3) Issue resolution results over four runs on DeepSWE with DeepSeek. Mean denotes the mean resolution rate (\%) across the four runs, with 95\% confidence intervals. Pass@4 and All@4 denote the proportions (\%) of issues resolved in at least one run and in all four runs, respectively.}
\label{tab:stability}
\scriptsize
\setlength{\tabcolsep}{5pt}
\begin{tabular}{lcrr}
\toprule
Method & Mean & Pass@4 & All@4 \\
\midrule
mini-SWE-agent & 45.6 $\pm$ 2.50 & 77.9 & 17.7 \\
Trace2Skill & 49.3 $\pm$ 2.69 & 81.4 & 21.2 \\
\midrule
\ours & \textbf{58.6 $\pm$ 2.49} & \textbf{89.4} & \textbf{32.7} \\
\bottomrule
\end{tabular}
\vspace{-10pt}
\end{wraptable}
RQ3 examines whether the effectiveness of \ours remains stable across repeated runs. In this RQ, we focus on DeepSWE, whose long-horizon tasks make repeated-run stability particularly relevant, and compare \ours with the raw mini-SWE-agent without learned skills and with Trace2Skill, the strongest baseline on DeepSWE in RQ1. Given the computational cost of repeated evaluation and our limited budget, we use DeepSeek as the backbone LLM. Following the DeepSWE evaluation protocol, we run each method four times on all tasks and report the mean resolution rate with its 95\% confidence interval. We further report Pass@4 and All@4, the proportions of issues resolved in at least one and in all four runs, respectively.

Table~\ref{tab:stability} reports the results. \ours{} maintains a clear advantage across repeated runs, achieving a mean resolution rate of 58.6\%, 13.0 and 9.3 percentage points higher than mini-SWE-agent and Trace2Skill, respectively. Its 95\% confidence interval does not overlap with either baseline. \ours{} also achieves the highest Pass@4 (89.4\%) and All@4 (32.7\%), exceeding Trace2Skill by 8.0 and 11.5 points, respectively. These results show that the gains observed in RQ1 persist across repeated runs rather than arising from a favorable single execution.

\subsection{RQ4: Ablation Study}
\label{sec:ablation}

RQ4 evaluates the key design choices in \ours. As in RQ3, we conduct the ablation study on DeepSWE with DeepSeek as the backbone LLM. We construct five variants spanning rule discovery, within-repository verification, cross-repository generalization, and rule selection. Each variant modifies only the designated component while keeping the rest of the pipeline unchanged: (1) \textbf{w/ Trace2Skill} replaces the candidate rules discovered in Stage~1 with rules generated by Trace2Skill; (2) \textbf{w/o within-repo} removes within-repository rule verification and directly passes rules to cross-repository consolidation; (3) \textbf{w/o cross-repo} removes cross-repository consolidation and retains the independently verified repository-level rules; (4) \textbf{w/ all rules} provides the full rule set to the agent instead of selecting relevant rules in Stage~3; and (5) \textbf{w/o evidence} removes the historical evidence associated with each rule from the Stage~3 selection process.
\begin{wraptable}{r}{0.39\textwidth}
\vspace{-12pt}
\centering
\caption{(RQ4) Issue resolution rates (\%) of \ours and its ablation variants on DeepSWE with DeepSeek. }
\label{tab:ablation}
\scriptsize
\setlength{\tabcolsep}{5pt}
\begin{tabular}{lr}
\toprule
Method & \%Resolved \\
\midrule
\ours & \textbf{58.4} \\
\midrule
 w/ Trace2Skill & 54.9 \\
w/o within-repo & 53.1 \\
 w/o cross-repo & 51.3 \\
 w/ all rules & 48.7 \\
 w/o evidence & 55.8 \\
\bottomrule
\end{tabular}
\vspace{-10pt}
\end{wraptable}

Table~\ref{tab:ablation} reports the results. Each ablation variant degrades the performance of \ours, suggesting that its gains do not stem from a single component but from the combined contributions of rule discovery, verification, cross-repository generalization, and rule selection. The largest drop comes from replacing issue-adaptive selection with all rules (58.4\% $\rightarrow$ 48.7\%), followed by removing cross-repository generalization (51.3\%) and removing within-repository verification (53.1\%). These results highlight the importance of verifying rules and consolidating behaviors recurring across repositories before inference and selecting only rules relevant to the current issue. Replacing divergence-guided discovery with Trace2Skill reduces the resolution rate to 54.9\%, suggesting that the source of candidate rules also matters, while removing rule evidence lowers performance to 55.8\%, suggesting that historical support helps the selector prioritize useful guidance.

\section{Related Work}
\label{app:related}

\paragraph{Software engineering agents.} Software engineering agents primarily target repository-level issue resolution. SWE-bench~\citep{jimenez2024swebench} established this setting as a standard benchmark and spurred rapid progress in systems such as SWE-agent~\citep{yang2024sweagent}, OpenHands~\citep{wang2025openhands}, AutoCodeRover~\citep{zhang2024autocoderover}, and Agentless~\citep{xia2025agentless}. mini-SWE-agent provides a lightweight, shell-based implementation of this workflow and has become a widely used open-source research baseline, with several recent agents building on it~\citep{guo2026swe,xia2025liveswe}. Subsequent work has further improved issue-resolution performance through stronger search and planning, multi-agent collaboration, and test-time scaling~\citep{tao2024magis,zhang2025dei,gao2025trae,ehrlich2025codemonkeys}. Our work studies skill learning in the same setting using mini-SWE-agent as the underlying agent, while keeping the agent itself unchanged so that performance differences reflect the skills rather than changes to the agent architecture.

\paragraph{Skill learning for agents.}
Skills provide agents with reusable task guidance at inference time. Their effectiveness, however, depends strongly on how they are constructed~\citep{han2026sweskillsbench}.
A growing body of work distills reusable skills from agent trajectories, including executable programs, natural-language experience, workflows, and skill documents~\citep{wang2023voyager,zheng2025skillweaver,zhao2024expel,ouyang2025reasoningbank,wang2024awm}. Among recent skill-learning methods, SkillOpt~\citep{yang2026skillopt} optimizes skills against validation performance, while Trace2Skill~\citep{ni2026trace2skill} consolidates recurring trajectory-local lessons into transferable skills. In software engineering specifically, STAIR~\citep{xu2026stair} abstracts repair trajectories into reusable plans for new issues. \ours differs by explicitly verifying candidate rules on trajectories beyond those from which they are discovered and retaining them as transferable only when the same behavior recurs across repositories.

\section{Conclusion}
This paper studies how to learn software engineering skills that remain useful beyond the repositories from which they are distilled. We introduce \ours, a trajectory-based skill learning approach that contrasts successful and failed attempts to discover candidate rules, verifies their prescribed behaviors on independent trajectories through executable predicates, and retains only rules independently verified across repositories as transferable guidance. Evaluated on SWE-bench Pro and DeepSWE with three backbone LLMs, \ours consistently achieves the highest issue resolution rate in all six benchmark--LLM combinations, compared with both the underlying agent without learned skills and recent skill-learning baselines. These results highlight the value of going beyond extracting recurring behaviors from past experience by testing candidate behaviors against broader trajectory evidence and validating them beyond their original repository context.

\subsection*{AI use statement}
Generative AI tools are used in this work in three ways. First, LLMs are an integral part of the proposed method and experimental pipeline. Second, we use generative AI tools to assist with code implementation. Third, we use generative AI tools to improve the clarity, grammar, and presentation of the manuscript.

All AI-assisted code is reviewed by the authors and tested as part of the experimental pipeline. All AI-assisted writing is manually reviewed, revised, and verified by the authors. The authors design the research methodology, conduct and validate the experiments, analyze the results, and determine the scientific claims and conclusions. We take responsibility for the final content of this work, including all text, code, claims, and artifacts produced with the aid of generative AI.

\subsection*{Reproducibility statement}
To support reproducibility, we have released the source code, data, and skill corpus used in this work at \url{https://github.com/XREPOSKILL/XRepoSkill}.

\bibliography{references}

@inproceedings{jimenez2024swebench,
  title={{SWE-bench}: Can Language Models Resolve Real-World {GitHub} Issues?},
  author={Jimenez, Carlos E. and Yang, John and Wettig, Alexander and Yao, Shunyu and Pei, Kexin and Press, Ofir and Narasimhan, Karthik},
  booktitle={International Conference on Learning Representations (ICLR)},
  year={2024}
}

@inproceedings{yang2024sweagent,
  title={{SWE-agent}: Agent-Computer Interfaces Enable Automated Software Engineering},
  author={Yang, John and Jimenez, Carlos E. and Wettig, Alexander and Lieret, Kilian and Yao, Shunyu and Narasimhan, Karthik and Press, Ofir},
  booktitle={Advances in Neural Information Processing Systems (NeurIPS)},
  year={2024}
}

@inproceedings{tripathy2026swenergy,
  title     = {{SWEnergy}: An Empirical Study on Energy Efficiency in Agentic Issue Resolution Frameworks with {SLMs}},
  author    = {Tripathy, Arihant and Harshit, Ch Pavan and Vaidhyanathan, Karthik},
  booktitle = {Proceedings of the 2026 International Workshop on Agentic Engineering (AGENT '26)},
  pages     = {104--111},
  publisher = {ACM},
  year      = {2026},
  doi       = {10.1145/3786167.3788406}
}

@article{guo2026swe,
  title   = {{SWE-Doctor}: Guiding Software Engineering Agents with Runtime Diagnosis from Multi-Faceted Bug Reproduction Tests},
  author  = {Guo, Yaoqi and Liu, Yang and Zhang, Jie M. and Ma, Yun and Lou, Yiling and Chen, Zhenpeng},
  journal = {arXiv preprint arXiv:2607.00990},
  year    = {2026}
}

@inproceedings{lee2026gistify,
  title     = {{Gistify}: Codebase-Level Understanding via Runtime Execution},
  author    = {Lee, Hyunji and Kim, Minseon and Singh, Chinmay and Pereira, Matheus and Sonwane, Atharv and White, Isadora and Stengel-Eskin, Elias and Bansal, Mohit and Shi, Zhengyan and Sordoni, Alessandro and C{\^o}t{\'e}, Marc-Alexandre and Yuan, Xingdi and Caccia, Lucas},
  booktitle = {International Conference on Learning Representations (ICLR)},
  year      = {2026}
}

@misc{swebench2026leaderboard,
  title={{SWE-bench} Leaderboards},
  author={{SWE-bench}},
  howpublished={\url{https://www.swebench.com/}},
  year={2026},
  note={Accessed September 18, 2026}
}

@misc{anthropic2026skills,
  author       = {{Anthropic}},
  title        = {Agent {Skills}},
  howpublished = {\url{https://platform.claude.com/docs/en/agents-and-tools/agent-skills/overview}},
  year         = {2026},
  note         = {Accessed September 18, 2026}
}

@article{mantel1959statistical,
  title={Statistical Aspects of the Analysis of Data from Retrospective Studies of Disease},
  author={Mantel, Nathan and Haenszel, William},
  journal={Journal of the National Cancer Institute},
  volume={22},
  number={4},
  pages={719--748},
  year={1959}
}

@article{sato1989variance,
  author  = {Sato, Tosiya and Greenland, Sander and Robins, James M.},
  title   = {On the Variance Estimator for the {Mantel--Haenszel} Risk Difference},
  journal = {Biometrics},
  volume  = {45},
  number  = {4},
  pages   = {1323--1324},
  year    = {1989}
}

@article{cochran1954combination,
  title={The Combination of Estimates from Different Experiments},
  author={Cochran, William G.},
  journal={Biometrics},
  volume={10},
  number={1},
  pages={101--129},
  year={1954}
}

@article{han2026sweskillsbench,
  title={{SWE-Skills-Bench}: Do Agent Skills Actually Help in Real-World Software Engineering?},
  author={Han, Tingxu and Zhang, Yi and Song, Wei and Fang, Chunrong and Chen, Zhenyu and Sun, Youcheng and Hu, Lijie},
  journal={arXiv preprint arXiv:2603.15401},
  year={2026}
}

@article{ni2026trace2skill,
  title={{Trace2Skill}: Distill Trajectory-Local Lessons into Transferable Agent Skills},
  author={Ni, Jingwei and Liu, Yihao and Liu, Xinpeng and Sun, Yutao and Zhou, Mengyu and Cheng, Pengyu and Wang, Dexin and Zhao, Erchao and Jiang, Xiaoxi and Jiang, Guanjun},
  journal={arXiv preprint arXiv:2603.25158},
  year={2026}
}

@article{yang2026skillopt,
  title={{SkillOpt}: Executive Strategy for Self-Evolving Agent Skills},
  author={Yang, Yifan and Gong, Ziyang and Huang, Weiquan and Yang, Qihao and Zhou, Ziwei and Huang, Zisu and Li, Yan and Gao, Xuemei and Dai, Qi and Liu, Bei and Qiu, Kai and Yang, Yuqing and Chen, Dongdong and Yang, Xue and Luo, Chong},
  journal={arXiv preprint arXiv:2605.23904},
  year={2026}
}

@article{xu2026stair,
  title={Reusing Past Repairs Through Hierarchical Trajectory Abstraction for Coding Agents},
  author={Xu, Yisen and Zhou, Jiayuan and Pan, Ruiqi and Chen, Tse-Hsun},
  journal={arXiv preprint arXiv:2607.29658},
  year={2026}
}

@inproceedings{deng2025swebenchpro,
  title     = {{SWE-Bench Pro}: Can {AI} Agents Solve Long-Horizon Software Engineering Tasks?},
  author    = {Deng, Xiang and Da, Jeff and Pan, Edwin and He, Yannis Yiming and Ide, Charles and Garg, Kanak and Lauffer, Niklas and Park, Andrew and Rane, Chetan and Sampath, Karmini and Krishnan, Maya and Kundurthy, Srivatsa and Hendryx, Sean and Wang, Zifan and Zhang, Chen Bo Calvin and Jacobson, Noah and Liu, Bing and Kenstler, Brad},
  booktitle = {Proceedings of the 43rd International Conference on Machine Learning (ICML)},
  year      = {2026}
}

@article{datacurve2026deepswe,
  title={{DeepSWE}: Measuring Frontier Coding Agents on Original, Long-Horizon Engineering Tasks},
  author={Huang, Wenqi and Lee, Charley and Tng, Leonard and Ge, Serena},
  journal={arXiv preprint arXiv:2607.07946},
  year={2026}
}

@inproceedings{wang2025openhands,
  title={{OpenHands}: An Open Platform for {AI} Software Developers as Generalist Agents},
  author={Wang, Xingyao and Li, Boxuan and Song, Yufan and Xu, Frank F. and Tang, Xiangru and Zhuge, Mingchen and Pan, Jiayi and Song, Yueqi and Li, Bowen and Singh, Jaskirat and Tran, Hoang H. and Li, Fuqiang and Ma, Ren and Zheng, Mingzhang and Qian, Bill and Shao, Yanjun and Muennighoff, Niklas and Zhang, Yizhe and Hui, Binyuan and Lin, Junyang and Brennan, Robert and Peng, Hao and Ji, Heng and Neubig, Graham},
  booktitle={International Conference on Learning Representations (ICLR)},
  year={2025}
}

@inproceedings{zhang2024autocoderover,
  title={{AutoCodeRover}: Autonomous Program Improvement},
  author={Zhang, Yuntong and Ruan, Haifeng and Fan, Zhiyu and Roychoudhury, Abhik},
  booktitle={Proceedings of the 33rd ACM SIGSOFT International Symposium on Software Testing and Analysis (ISSTA)},
  pages={1592--1604},
  doi={10.1145/3650212.3680384},
  year={2024}
}

@article{xia2025agentless,
  title={Demystifying {LLM}-based Software Engineering Agents},
  author={Xia, Chunqiu Steven and Deng, Yinlin and Dunn, Soren and Zhang, Lingming},
  journal={Proceedings of the ACM on Software Engineering (PACMSE)},
  volume={2},
  number={FSE},
  pages={801--824},
  doi={10.1145/3715754},
  year={2025}
}

@inproceedings{tao2024magis,
  title={{MAGIS}: {LLM}-Based Multi-Agent Framework for {GitHub} Issue Resolution},
  author={Tao, Wei and Zhou, Yucheng and Wang, Yanlin and Zhang, Wenqiang and Zhang, Hongyu and Cheng, Yu},
  booktitle={Advances in Neural Information Processing Systems (NeurIPS)},
  year={2024}
}

@inproceedings{zhang2025dei,
  title={Diversity Empowers Intelligence: Integrating Expertise of Software Engineering Agents},
  author={Zhang, Kexun and Yao, Weiran and Liu, Zuxin and Feng, Yihao and Liu, Zhiwei and Murthy, Rithesh and Lan, Tian and Li, Lei and Lou, Renze and Xu, Jiacheng and Pang, Bo and Zhou, Yingbo and Heinecke, Shelby and Savarese, Silvio and Wang, Huan and Xiong, Caiming},
  booktitle={International Conference on Learning Representations (ICLR)},
  year={2025}
}

@article{gao2025trae,
  title={{Trae Agent}: An {LLM}-based Agent for Software Engineering with Test-time Scaling},
  author={Gao, Pengfei and Tian, Zhao and Meng, Xiangxin and Wang, Xinchen and Hu, Ruida and Xiao, Yuanan and Liu, Yizhou and Zhang, Zhao and Chen, Junjie and Gao, Cuiyun and Lin, Yun and Xiong, Yingfei and Peng, Chao and Liu, Xia},
  journal={arXiv preprint arXiv:2507.23370},
  year={2025}
}

@article{ehrlich2025codemonkeys,
  title={{CodeMonkeys}: Scaling Test-Time Compute for Software Engineering},
  author={Ehrlich, Ryan and Brown, Bradley and Juravsky, Jordan and Clark, Ronald and R{\'e}, Christopher and Mirhoseini, Azalia},
  journal={arXiv preprint arXiv:2501.14723},
  year={2025}
}

@article{xia2025liveswe,
  title={{Live-SWE-agent}: Can Software Engineering Agents Self-Evolve on the Fly?},
  author={Xia, Chunqiu Steven and Wang, Zhe and Yang, Yan and Wei, Yuxiang and Zhang, Lingming},
  journal={arXiv preprint arXiv:2511.13646},
  year={2025}
}

@article{wang2023voyager,
  title   = {Voyager: An Open-Ended Embodied Agent with Large Language Models},
  author  = {Wang, Guanzhi and Xie, Yuqi and Jiang, Yunfan and Mandlekar, Ajay and Xiao, Chaowei and Zhu, Yuke and Fan, Linxi and Anandkumar, Anima},
  journal = {Transactions on Machine Learning Research},
  year    = {2024},
}

@inproceedings{zhao2024expel,
  title={{ExpeL}: {LLM} Agents Are Experiential Learners},
  author={Zhao, Andrew and Huang, Daniel and Xu, Quentin and Lin, Matthieu and Liu, Yong-Jin and Huang, Gao},
  booktitle={Proceedings of the AAAI Conference on Artificial Intelligence (AAAI)},
  year={2024}
}

@inproceedings{ouyang2025reasoningbank,
  title     = {{ReasoningBank}: Scaling Agent Self-Evolving with Reasoning Memory},
  author    = {Ouyang, Siru and Yan, Jun and Hsu, I-Hung and Chen, Yanfei and Jiang, Ke and Wang, Zifeng and Han, Rujun and Le, Long and Daruki, Samira and Tang, Xiangru and Tirumalashetty, Vishy and Lee, George and Rofouei, Mahsan and Lin, Hangfei and Han, Jiawei and Lee, Chen-Yu and Pfister, Tomas},
  booktitle = {International Conference on Learning Representations (ICLR)},
  year      = {2026}
}

@inproceedings{wang2024awm,
  title={Agent Workflow Memory},
  author={Wang, Zora Zhiruo and Mao, Jiayuan and Fried, Daniel and Neubig, Graham},
  booktitle={Proceedings of the 42nd International Conference on Machine Learning (ICML)},
  volume={267},
  series={PMLR},
  pages={63897--63911},
  year={2025}
}

@article{zheng2025skillweaver,
  title={{SkillWeaver}: Web Agents can Self-Improve by Discovering and Honing Skills},
  author={Zheng, Boyuan and Fatemi, Michael Y. and Jin, Xiaolong and Wang, Zora Zhiruo and Gandhi, Apurva and Song, Yueqi and Gu, Yu and Srinivasa, Jayanth and Liu, Gaowen and Neubig, Graham and Su, Yu},
  journal={arXiv preprint arXiv:2504.07079},
  year={2025}
}

@article{defays1977efficient,
  title={An efficient algorithm for a complete link method},
  author={Defays, Daniel},
  journal={The Computer Journal},
  volume={20},
  number={4},
  pages={364--366},
  year={1977}
}

@article{yule1912methods,
  title={On the methods of measuring association between two attributes},
  author={Yule, G. Udny},
  journal={Journal of the Royal Statistical Society},
  volume={75},
  number={6},
  pages={579--652},
  year={1912}
}
\bibliographystyle{arxiv}

\appendix
\section{Implementation Details}
\label{app:impl}

Section~\ref{sec:method} describes the design of \ours. This appendix lists, stage by stage, the parameters, prompt constraints, and formats with which the design is implemented, using the terms of Section~\ref{sec:method}. Every LLM call in Stage~1 and Stage~2 is made to DeepSeek-V4-Pro-0813, the LLM that \ours uses for skill learning in Section~\ref{sec:eval}; we refer to it below as the skill-learning LLM. The two calls of Stage~3 are made to the backbone LLM of the agent $A$.

\subsection{Stage 1: Divergence-Guided Rule Discovery}
\label{app:impl:fork}

\paragraph{Pairing.}
Backbone LLMs are ranked by their resolution rate on the trajectory pool. On each issue with both successful and failed trajectories, every failed trajectory is combined with every successful trajectory from a backbone LLM with a higher resolution rate, the combinations are ranked by the gap in resolution rate between the two LLMs, and the top four become pairs. A check before pairing asserts that no issue of the pool appears in either evaluation benchmark.

\paragraph{Action abstraction.}
Before a tool call is mapped to an abstract action, compound commands joined by \texttt{\&\&} or \texttt{;} are split into one command each, and leading \texttt{cd}, environment variable assignments, and \texttt{timeout} prefixes are stripped. A bare \texttt{cd}, \texttt{pwd}, or \texttt{export} yields no action. The type of the action is then decided by the program that leads the command. A read prints part of a file with \texttt{cat}, \texttt{head}, \texttt{tail}, \texttt{less}, \texttt{nl}, \texttt{sed -n}, or \texttt{awk}; its target is the first file path in the command. A search runs \texttt{grep}, \texttt{rg}, \texttt{ack}, \texttt{find}, \texttt{ls}, \texttt{which}, \texttt{locate}, or \texttt{tree}; its target is the quoted pattern, or the first non-option argument when nothing is quoted. An edit runs \texttt{sed -i}, redirects the output of \texttt{cat}, \texttt{echo}, \texttt{printf}, or \texttt{tee} into a source or configuration file, or runs \texttt{patch} or \texttt{git apply}; its target is the written file. A test run invokes \texttt{pytest}, \texttt{tox}, \texttt{python -m pytest}, \texttt{python -m unittest}, or a \texttt{runtests.py} script; its target is the first argument containing ``test''. A script run invokes \texttt{python} or \texttt{ipython} for anything other than a test run; its target is the first file path. An install runs \texttt{pip}, \texttt{conda}, or \texttt{apt}; its target is the package name. A git action is any other \texttt{git} subcommand, with the subcommand as target. A submission is the final-output marker of mini-SWE-agent. Every remaining command is of type other, with an empty target. A regular expression in a predicate is matched case-insensitively against the full command text before splitting and stripping.

\paragraph{Divergence localization.}
The shared prefix is extended across matching blocks of the two action sequences. A gap of unmatched actions between two matching blocks is tolerated only if it holds at most three actions on each side and none of them is an edit, a test run, a script run, an install, or a submission. The text handed to the skill-learning LLM contains the shared prefix as a compact list of actions, the two trajectory suffixes from the divergence point onward within a fixed character budget, a compact list of the actions that follow the budget, and the position of every command in its trajectory, so that a prompt can cite a command by position.

\paragraph{Candidate rule generation.}
Each pair is read by three prompts, one per reference point of Section~\ref{sec:method:fork}, and each prompt sees the text described above and nothing else from the trajectories. The first prompt reads the failed suffix together with the patch that the successful trajectory submitted and reports the earliest step that the failed trajectory skipped. The second reads the successful suffix only and reports up to three steps taken before its first edit. The third reads both suffixes and reports the earliest step that only the successful trajectory took, citing one command position on each side. The first and third prompts return one candidate rule each and the second returns one to three, so each pair yields three to five candidate rules. Each candidate rule is returned as a JSON object with a paragraph of at most 1,000 characters, a category, one of lookup, verification, replay, or search recipe, the trajectory identifiers and command positions it cites as evidence, and its executable predicate. A candidate rule that states a fact about the code instead of prescribing an action carries no predicate and must name at least one real file path or symbol; such rules cannot be verified in Stage~2 and never enter the transferable skill.

\paragraph{Rejection.}
A candidate rule is rejected if its paragraph exceeds the length limit, if it prescribes an action and contains a file path followed by a line number or three or more rare symbols such as double-underscore method names, if its predicate does not parse, or if its predicate is true on fewer than 10\% or more than 90\% of the trajectories of the pool. A rejected rule is re-prompted once with the reason and dropped if it fails again.

\subsection{Stage 2: Rule Verification and Cross-Repository Generalization}
\label{app:impl:distill}

\paragraph{Within-repository merging.}
Within a repository, candidate rules of the same category are handed to the skill-learning LLM in batches, and each batch is merged into between 3 and 10 rules; rules with the same title from different batches are merged again. The number of candidate rules merged into a rule is its support, and the predicate is also selected by the LLM. The prompt requires that exact commands, search patterns, and paths present in the candidate rules be kept verbatim, and rejects a merged rule reduced to advice such as ``reproduce first''. A screen by the skill-learning LLM drops a merged rule only if it neither prescribes an action nor states a fact anchored to a file path or symbol, or if it contradicts conventions already recorded for its repository. An n-gram audit flags any rule whose text shares six or more consecutive tokens with the reference patch of one of its source issues.

\paragraph{Resolution gain and z score.}
The weight $w_j$ in ~\eqref{eq:gain} is largest when both groups of issue $j$ are large and is zero when every trajectory of the issue falls in one group. The standard error of the gain is the estimator of \citet{sato1989variance}. Within each repository, rules are ordered by z score. A rule whose predicate takes the same value on every trajectory of its repository receives no gain; such rules and rules that state facts are placed at the tail of the order, ordered by support.

\paragraph{Cross-repository generalization.}
The correlation $\phi$ is computed only for verified rules. The cutoff on $\phi$ is read from the histogram of $\phi$ over all pairs of rules, which has a spike near one and a broader mass at lower values; the cutoff is the local minimum between the two. When the rules of a cluster are rewritten into a transferable rule, at most one verbatim example command per repository is kept, and rules that give different advice are listed as outliers. In ~\eqref{eq:pool}, the standard error $s_k$ of each merged rule is its gain divided by its z score, and the support of the transferable rule is the sum of the supports of the rules it merges.

\paragraph{Skill file.}
The skill $S$ is written as one Markdown file. Each transferable rule is one bullet holding its title, a tag, and its body, and the bullets are ordered by z score. The tag holds the gain of the rule and its z score, for example ``(gain +5.7pp, z +15.59)''. Rules whose cluster came from a single repository do not enter $S$.

\subsection{Stage 3: Issue-Adaptive Rule Selection}
\label{app:impl:inject}

\paragraph{Rule selection.}
Selection uses two calls to the backbone LLM of $A$. The first call receives the issue description, cut to 6,000 characters, and the titles of all rules, and returns a summary of the bug in two to three sentences and a list of the kinds of guidance that would help. The second call receives the issue description, that summary, and every rule with its identifier, tag, title, and body in z score order, and returns at most three rule identifiers with a one-line reason each. The system prompt of the second call asks for the rules most likely to help an engineer fix the bug. Identifiers that do not exist are discarded, and the selection may be empty. In our experiments, the rules shown to the two calls are the 36 transferable rules of $S$; Appendix~\ref{app:selection} reports which of them were chosen.

\paragraph{Skill-guided resolution.}
The selected rules are placed above the issue description in the first user message of mini-SWE-agent, with the framing that they are guidance on how to investigate and acquire facts at runtime and that each step must be verified against the actual code before acting. When the selection is empty, the agent runs without added text. The agent makes no further calls for rules during its loop, and the summary, the reasons, and the tags never enter its prompt.

\section{Details of the Evaluation}
\label{app:eval}

\subsection{Datasets}
\label{app:eval:data}

The trajectory pool of Section~\ref{sec:eval} is downloaded from the public SWE-bench experiments repository. Its 500 issues come from 12 Python repositories, which are the repositories within which Stage~2 verifies rules and across which it merges them. The 200 SWE-bench Pro instances of the main results cover all 11 repositories of the public set of the benchmark.
Each DeepSWE task specifies a change to a repository in natural language and is graded by hidden tests.

\subsection{Evaluation Protocol}
\label{app:eval:protocol}

All methods run the same mini-SWE-agent with the same system prompt, tool interface, and step and cost limits, and each method changes only the text placed before the issue description in the first user message. The skills of the three baselines are learned from the same trajectory pool with the skill-learning LLM of Appendix~\ref{app:impl}. The per-issue plan calls of STAIR and the two selection calls of \ours use the backbone LLM that runs the repair. Except in the stability study of Section~\ref{sec:stability}, each issue is run once per method and backbone LLM, and the reported resolution rates come from that single run.

SWE-bench Pro issues are mined from public repositories, and the commit that fixed each issue is present in the public history of the repository, so an agent that can read that history or reach the network can copy the fix. We therefore build the container of every SWE-bench Pro issue in three steps. First, the repository is checked out at the base commit that the issue specifies. Second, the \texttt{.git} directory is deleted after the checkout, so \texttt{git log}, \texttt{git diff}, \texttt{git show}, and every other command that reads the history fail, and the agent sees only the files of the working tree. Third, the container is started without network access, so \texttt{git fetch}, \texttt{curl}, \texttt{pip install}, and any other command that reaches outside the container fail. The official evaluation images keep the history and the network, so resolution rates measured in them are not directly comparable to ours. Every method and backbone LLM in the paper runs in the same containers, so the comparison between methods is unaffected. DeepSWE tasks are authored for the benchmark rather than mined from public repositories, so no upstream fix exists to be found, and we run them in the official environment.

\subsection{Cost Accounting}
\label{app:eval:cost}

The cost per issue reported in Section~\ref{sec:main} is computed from the token counts recorded in the trajectory of each run and in the records of the calls made before the repair loop starts, namely the two selection calls of \ours and the three plan calls of STAIR. Every call is priced with the list price of its provider, in US dollars per million tokens. Input tokens are split into cache hits and cache misses, and each part is billed at the rate the provider charges for it: 0.003 and 0.15 for DeepSeek at the off-peak rate, 0.02 and 0.20 for GPT, whose cache writes are billed at 0.25, and 0.0036 and 0.435 for MiMo. Output tokens are billed at 0.60, 1.20, and 0.87, respectively. Table~\ref{tab:cost} lists the resulting cost per issue for every method, backbone LLM, and benchmark; the averages over backbone LLMs are the numbers quoted in Section~\ref{sec:main}. The two selection calls of \ours cost \$0.01 per issue with DeepSeek and GPT and \$0.03 with MiMo, and the rest of its cost comes from the repair loop.

\begin{table}[h]
\centering
\caption{Cost per issue in US dollars for every method, backbone LLM, and benchmark of the main results. Avg. is the mean over the three backbone LLMs on the same benchmark.}
\label{tab:cost}
\setlength{\tabcolsep}{4pt}
\small
\begin{tabular}{lrrrrrrrr}
\toprule
& \multicolumn{4}{c}{SWE-bench Pro}
& \multicolumn{4}{c}{DeepSWE} \\
\cmidrule(lr){2-5}\cmidrule(lr){6-9}
Method
& \multicolumn{1}{c}{DeepSeek}
& \multicolumn{1}{c}{GPT}
& \multicolumn{1}{c}{MiMo}
& \multicolumn{1}{c}{Avg.}
& \multicolumn{1}{c}{DeepSeek}
& \multicolumn{1}{c}{GPT}
& \multicolumn{1}{c}{MiMo}
& \multicolumn{1}{c}{Avg.} \\
\midrule
mini-SWE-agent & 0.133 & 0.089 & 0.034 & 0.086 & 0.150 & 0.390 & 0.128 & 0.223 \\
\midrule
SkillOpt       & 0.167 & 0.148 & 0.041 & 0.119 & 0.135 & 0.407 & 0.125 & 0.222 \\
STAIR          & 0.140 & 0.098 & 0.052 & 0.097 & 0.150 & 0.358 & 0.140 & 0.216 \\
Trace2Skill    & 0.153 & 0.117 & 0.036 & 0.102 & 0.139 & 0.274 & 0.124 & 0.179 \\
\midrule
\ours          & 0.156 & 0.110 & 0.065 & 0.110 & 0.161 & 0.426 & 0.152 & 0.246 \\
\bottomrule
\end{tabular}
\end{table}

\begin{figure}[t]
\centering
\includegraphics[width=\linewidth]{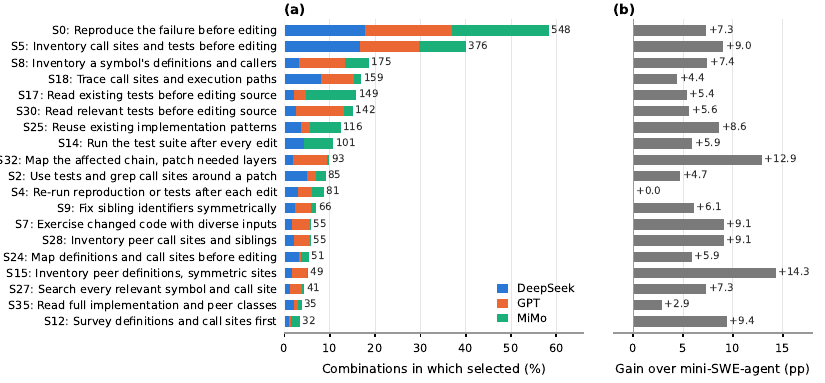}
\caption{The 19 transferable rules selected in at least 30 of the 939 combinations of an issue and a backbone LLM in the main results, ordered by selection count. (a) Share of combinations in which the rule was selected, stacked by backbone LLM; the number at the end of a bar is the count. (b) Resolution rate of \ours minus that of mini-SWE-agent, in percentage points, on the combinations in which the rule was selected.}
\label{fig:rule_selection}
\end{figure}

\section{Which Rules Are Selected, and with What Result}
\label{app:selection}

The main results place at most three rules in the prompt of each issue, and the second selection call of Stage~3 decides which. This section examines the selections made for the 939 combinations of an issue and a backbone LLM in the main results, recovered from the prompt of each run, and relates every transferable rule to the outcome on the combinations in which it was selected.

\paragraph{Selection concentrates on a few rules.}
Figure~\ref{fig:rule_selection}(a) shows the 19 transferable rules selected in at least 30 combinations. S0, reproduce the reported failure before touching source, was selected in 58.4\% of the combinations and S5, inventory call sites and tests before modifying shared behavior, in 40.0\%. The five most selected rules account for half of all selections and the top 11 for 80\%. Of the other 17 rules, four were never selected and eight were selected fewer than ten times, so a third of the skill went almost unused. Part of this tail consists of rules that prescribe the same action as a frequent one in different words, for example, to read the tests before editing. Cross-repository generalization kept them apart because their predicates are true on different trajectories, and the selection call spreads its choices over them.

Which rules are selected depends on the backbone LLM that selects. MiMo selected S17 in 33\% of its combinations against 6\% for DeepSeek and 8\% for GPT. GPT selected S30 in 32\% of its combinations and never selected S14. DeepSeek selected S5 in 50\% of its combinations. The Spearman rank correlation of selection counts between two backbone LLMs lies between 0.68 and 0.81, whereas between the two benchmarks, with backbone LLMs pooled, it is 0.93.

\paragraph{The agent resolves more issues when a rule is selected.}
Figure~\ref{fig:rule_selection}(b) shows, for the same 19 rules, the resolution rate of \ours minus that of mini-SWE-agent on the combinations in which the rule was selected. The difference is positive for 18 rules and zero for one, S4, and ranges from 2.9 to 14.3 percentage points; the two most selected rules, S0 and S5, gain 7.3 and 9.0 points on 548 and 376 combinations. Because up to three rules are placed in the prompt together and the combinations are chosen by the selection call rather than at random, a gain belongs to the selected set rather than to one rule.

\end{document}